\documentclass[a4paper,11pt,openright,twoside]{article}

\usepackage[english]{babel}
\usepackage[utf8x]{inputenc}

\usepackage[T1]{fontenc}

\usepackage{dsfont}

\usepackage{pdfpages}

\usepackage[toc,page]{appendix}
\usepackage{amssymb}
\usepackage{amsmath}
\usepackage{listings}
\usepackage{listingsutf8}
\usepackage{array}

\usepackage{arydshln}

\usepackage{lscape}
\usepackage{booktabs}

\usepackage{blindtext}

\usepackage{afterpage}
\usepackage{multirow}
\usepackage{longtable}
\usepackage{graphicx}
\usepackage{caption}
\usepackage{grffile}
\usepackage{float}

\usepackage[top=2cm,bottom=2cm,left=2cm,right=2cm]{geometry}

\usepackage{anyfontsize}

\usepackage{fancyhdr}

\usepackage{chngpage}

\usepackage{arydshln}

 \usepackage[flushleft]{threeparttable}
 
\DeclareMathOperator*{\argmin}{arg\,min}

\usepackage{multicol}

\usepackage{titlesec}
\titleformat{\chapter}[display]
{\normalfont\Large\filcenter\sffamily}
{\rule[1mm]{15mm}{1.5mm} \hspace{4mm} \LARGE{\chaptertitlename} \thechapter \hspace{5mm} \rule[1mm]{15mm}{1.5mm}}
{1pc}
{
\vspace{1pc}%
\Huge}

\usepackage{hyperref}

\usepackage{ulem}

\usepackage{amsbsy}

\begin{document}

\title{Bayesian dose-regimen assessment in early phase oncology incorporating pharmacokinetics and pharmacodynamics}

\date{}

\maketitle

\vspace{-2cm}

\begin{center}
Emma Gerard$^{1,2,3}$, Sarah Zohar$^{1,*}$, Hoai-Thu Thai$^{4}$, Christelle Lorenzato$^{2}$, \\
Marie-Karelle Riviere$^{3\ddagger}$ and Moreno Ursino$^{1,5\ddagger}$
\end{center}

\begin{center}
\footnotesize{$^{1}$ Centre de Recherche des Cordeliers, Sorbonne Universit\'e, Inserm, Universit\'e de Paris, F-75006, Paris, France} \\
\footnotesize{$^{2}$ Oncology biostatistics, Biostatistics and Programming department, Sanofi R\&D, Vitry-sur-Seine, France} \\
\footnotesize{$^{3}$ Statistical Methodology Group, Biostatistics and Programming department, Sanofi R\&D, Chilly-Mazarin, France} \\
\footnotesize{$^{4}$ Translation Disease Modeling, Digital and Data Science, Sanofi R\&D, Chilly-Mazarin, France} \\
\footnotesize{$^{5}$ F-CRIN PARTNERS platform, AP-HP, Universit\'e de Paris, Paris, France}\\
\footnotesize{$^{\ddagger}$ Co-last authors}\\
\footnotesize{$^{*}$ Email: sarah.zohar@inserm.fr}
\end{center}

\vspace{1cm}

\begin{footnotesize}
\begin{center}
\textbf{Abstract}
\end{center}
Phase I dose-finding trials in oncology seek to find the maximum tolerated dose (MTD) of a drug under a specific schedule. Evaluating drug-schedules aims at improving treatment safety while maintaining efficacy. However, while we can reasonably assume that toxicity increases with the dose for cytotoxic drugs, the relationship between toxicity and multiple schedules remains elusive. We proposed a Bayesian dose-regimen assessment method (DRtox) using pharmacokinetics/pharmacodynamics (PK/PD) information to estimate the maximum tolerated dose-regimen (MTD-regimen), at the end of the dose-escalation stage of a trial to be recommended for the next phase. We modeled the binary toxicity via a PD endpoint and estimated the dose-regimen toxicity relationship through the integration of a dose-regimen PD model and a PD toxicity model. For the dose-regimen PD model, we considered nonlinear mixed-effects models, and for the PD toxicity model, we proposed the following two Bayesian approaches: a logistic model and a hierarchical model. We evaluated the operating characteristics of the DRtox through simulation studies under various scenarios. The results showed that our method outperforms traditional model-based designs demonstrating a higher percentage of correctly selecting the MTD-regimen. Moreover, the inclusion of PK/PD information in the DRtox helped provide more precise estimates for the entire dose-regimen toxicity curve; therefore the DRtox may recommend alternative untested regimens for expansion cohorts. The DRtox should be applied at the end of the dose-escalation stage of an ongoing trial for patients with relapsed or refractory acute myeloid leukemia (NCT03594955) once all toxicity and PK/PD data are collected.\\

\end{footnotesize}

\section{Introduction}
\label{s:intro}

Phase I dose-finding clinical trials in oncology seek to find the maximum tolerated dose (MTD) to obtain reliable information regarding the safety profile of a drug or a combination of drugs, pharmacokinetics, and the mechanism of action \cite{chevret_2006}\cite{crowley_2005}. In this phase, the endpoint is defined as the dose-limiting toxicity (DLT), which is mainly based on the National Cancer Institute (NCI) Common Toxicity Criteria for Adverse Events \cite{CTCAE_2017}. Usually, standard algorithm-based or model-based dose-escalation methods aim to find the MTD while considering the entire cycle dosing as a single administration \cite{storer_1989}\cite{oquigley_1990}. Most methods assume that toxicity increases with the dose; however, the estimation of the relationship between toxicity and multiple doses over a cycle remains elusive as we can observe nonlinear dose-response profiles \cite{musuamba_2017}\cite{bullock_2017}\cite{schmoor_schumacher_1992}. We assume that considering the complete cycle dosage could improve treatment safety while maintaining future potential efficacy.

To account for dosage repetition over the treatment cycle, some authors have considered either the dose-schedule or the dose-regimen relationship. The National Cancer Institute defines “schedule” as “A step-by-step plan of the treatment that a patient is going to receive […] It also includes the amount of time between courses of treatment and the total length of time of treatment.” Moreover, the NCI defines “regimen” as “A treatment plan that specifies the dosage, the schedule, and the duration of treatment”. Following these definitions, we considered the dose-regimen relationship, as it includes the dosage, the repetition scheme and the duration.

For some molecules, it has been observed that, in the same patient, starting a dose-regimen with a lower lead-in dose and increasing the dose step-by-step before reaching the steady-state dose can reduce the occurrence of acute toxicities \cite{chen_2019}. However, a dose-regimen starting with higher lead-in doses can increase the efficacy. 

Dose-finding trials can aim to study different dose-regimens with the same or different total cumulative dose to determinate the most appropriate regimen supported by PK/PD profiles. Several methodological papers have attempted to address the issue of prospective dose and schedule finding methods. Braun et al., Liu and Braun and Zhang and Braun proposed considering the time-to-toxicity rather than the usual binary outcome to optimize dose and schedule, as the timing of administration \cite{braun_2005}\cite{braun_2007}\cite{liu_braun_2009}\cite{zhang_braun_2013}. Wages et al. proposed considering dose-schedule finding as a 2-dimensional problem and extended the partial order continual reassessment method developed for combination trials \cite{wages_2014}. Other authors proposed dose-schedule-finding methods that jointly model toxicity and efficacy outcomes \cite{li_2008}\cite{thall_2013}\cite{guo_2016}. Lyu et al. proposed a hybrid design that is partially algorithm-based and partially model-based for sequences of doses over multiple cycles when few doses are under study \cite{lyu_2018}.

Only a few methods consider PK/PD data in the prospective dose-allocation design. Ursino et al. compared multiple methods that enable the use of PK measures in sequential Bayesian adaptive dose-finding designs, including a dose-AUC-toxicity model combining two models to recommend the dose \cite{ursino_2017}. Gunhan et al. proposed a Bayesian time-to-event pharmacokinetic adaptive model for multiple regimens using PK latent profiles to measure drug exposure \cite{gunhan_2018}. Our aim is to extend these propositions by modeling the dose-regimen toxicity relationship using PK/PD.

\section{Motivation}

This work was motivated by the ongoing first-in-human dose-escalation study of SAR440234 administered as a single agent to patients with relapsed or refractory acute myeloid leukemia, high-risk myelodysplastic syndrome or B-cell acute lymphoblastic leukemia (NCT03594955) \cite{NCT03594955}. SAR440234 is a novel bispecific T-cell engager antibody that activates and redirects cytotoxic T lymphocytes (CTLs) to enhance the CTL-mediated elimination of CD123-expressing tumor cells. CTL activation induces the release of inflammatory cytokines, which can potentially cause cytokine release syndrome (CRS). CRS is a systemic inflammatory response and among the most commonly observed toxicities of T-cell engaging bispecific antibodies, such as blinatumomab,  which is a bispecific anti-CD19/CD3 antibody \cite{Shimabukuro-Vornhagen_2018}. Several cytokines, such as IL6, IL10 and INF$\gamma$, are consistently found to be elevated in serum from patients with CRS. The association between the peak of cytokine and CRS has been evaluated by Teachey et al. \cite{teachey_2016}. It has been shown that repeating the dosing of the drug can decrease CRS, particularly when the first administration is divided into several steps progressively \cite{chen_2019}. Therefore, intra-patient dose-escalation with a dose-regimen consisting of lower initial doses followed by a higher maintenance dose was implemented in this study to reduce the occurrence of CRS \cite{boissel_2018}.

The aim of the trial was to find the MTD of SAR440234 using the 3+3 design as the dose-escalation design. However, the 3+3 design and more general dose-finding designs do not consider intra-patient escalation information to decrease PD toxicity outcomes (CRS); these designs transform the dose-regimen received by the patient into a single dose-level. This approach is inefficient for achieving the trial goal. 

\begin{figure}
\begin{center}
\centerline{\includegraphics[width=16cm]{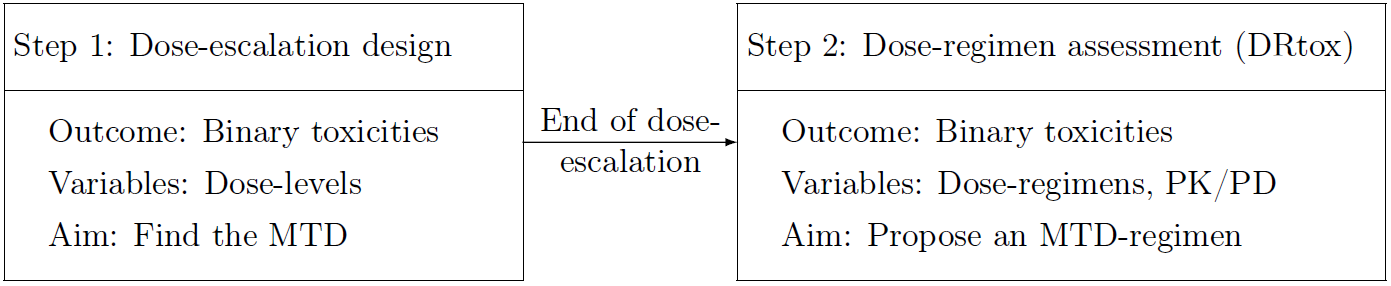}}
\end{center}
\caption{Trial scheme: the DRtox method is applied at the end of the dose-escalation stage of a phase I trial.
\label{trial_scheme}}
\end{figure}

In conclusion, we propose to model the binary toxicity endpoint (CRS) and the continuous PD response (cytokine profile) at the end of the trial, once all data have been collected, to characterize the dose-regimen toxicity relationship. This dose-regimen assessment method (DRtox) allows the determination of the maximum tolerated dose-regimen (MTD-regimen), as illustrated in Figure \ref{trial_scheme}.

\section{Model}
\label{model}

Let $\boldsymbol{\mathcal{D}}=\{d_1,...,d_L\}$ be the set of doses that can be administered to patients, where $d_{l}<d_{l+1}$. Let $\boldsymbol{\mathcal{S}}=\{\boldsymbol{S_1},...,\boldsymbol{S_K}\}\subset \boldsymbol{\mathbb{S}}$ be the panel of dose-regimens to be studied in the trial. The dose-regimen $\boldsymbol{S_k} \in \boldsymbol{\mathcal{S}}$, where $k \in \{1,...,K\}$, is defined as the sequence of $J$ doses, $\boldsymbol{S_k}=(d_{k,1},d_{k,2},...,d_{k,J})$, administered at times $\boldsymbol{t}=(t_1,t_2,...,t_J)$, where $d_{k,j} \in \boldsymbol{\mathcal{D}}$ for $j \in \{1,...,J\}$. To simplify the notations, we assumed that all dose-regimens have the same number of drug administrations at the same times, but this assumption can be relaxed. Let $\boldsymbol{S_{k,j}}$ be the subregimen of $\boldsymbol{S_k}$ until the $j^\text{th}$ administration, $\boldsymbol{S_{k,j}}=(d_{k,1},d_{k,2},...,d_{k,j})$, for $j<J$. Let $n \in \mathbb{N}$ be the number of patients included in the trial. Let $Y_{i,j}$ be the binary toxicity response of patient $i$ observed exactly after the $j^\text{th}$ administration, and let $Y_i$ be his/her global toxicity response at the end of the administrations.

Let $\boldsymbol{\widetilde{s}_{i}}=(d_{i,1},d_{i,2},...,d_{i,J}) \in \boldsymbol{\mathcal{S}}$ be the dose-regimen planned for the $i^\text{th}$ patient. We assume that the drug administration is stopped if toxicity occurs; thus let $j_i$ denote the last administration of patient $i$. We denote the actual regimen received by patient $i$ as $\boldsymbol{s_{i}}=(d_{i,1},d_{i,2},...,d_{i,j_{i}}) \subset \boldsymbol{\widetilde{s}_{i}}$,  where $\boldsymbol{s_{i}}=\boldsymbol{\widetilde{s}_{i}}$ if no toxicity is observed. Let $\boldsymbol{s_{i,j}}$ be the subregimen until $j$ of $\boldsymbol{s_i}$, where $j \leq j_i$.

The aim is to estimate the MTD-regimen at the end of the trial, which is defined as the dose-regimen with the toxicity probability closest to the target toxicity rate $\delta_{T}$, i.e. the MTD-regimen is the regimen $\boldsymbol{S_{k^\star}}$, where $k^\star=\displaystyle\argmin_{k} \left | p_T(\boldsymbol{S_k})- \delta_{T} \right | $.

We assume that a PD endpoint extracted from the continuous PD profile of a biomarker related to toxicity plays an intermediate role in the dose-regimen toxicity relationship. We propose a dose-regimen assessment method (DRtox) in which the first model is built for the dose-regimen and the PD endpoint, and the second model is built for the PD endpoint and the toxicity response. Therefore, integrating both models links the dose-regimen to the toxicity response to find the MTD-regimen. In the following section, the structure of the PK/PD models is described, two approaches between the PD endpoint and toxicity response are proposed, as well as a practical method for their integration.

\subsection{Dose-regimen PD response model}

Let $C(t)$ be the continuous drug concentration and $E(t)$ be the continuous PD response related to toxicity measured at time $t$.
We assume that $C(t)$ and $E(t)$ can be modeled using nonlinear mixed-effects models as follows:
\begin{equation}
\left\{
	\begin{array}{l}
		C(t)=f^{(1)}\left(\boldsymbol{\theta_i^{(1)}},t\right)+g^{(1)}\left(\boldsymbol{\theta_i^{(1)}},t,\boldsymbol{\xi_1}\right)\epsilon^{(1)} \\
		E(t)=f^{(2)}\left(\boldsymbol{\theta_i^{(2)}},t\right)+g^{(2)}\left(\boldsymbol{\theta_i^{(2)}},t,\boldsymbol{\xi_2}\right)\epsilon^{(2)} \\
	\end{array}
\right.
\end{equation}
where $f^{(1)}$ and $f^{(2)}$ represent the structural models, which are usually solutions of differential equations based on biological knowledge. $\boldsymbol{\theta_i}=\left(\boldsymbol{\theta_i^{(1)}},\boldsymbol{\theta_i^{(2)}}\right)$ represents the $i$th patient's specific parameter vector, where usually, $\boldsymbol{\theta_i}=\boldsymbol{\mu} e^{\boldsymbol{\eta_i}}$, with $\boldsymbol{\mu}$ denoting the fixed effects vector, and $\boldsymbol{\eta_i}$ denoting the random effects vector defined as $\boldsymbol{\eta_i} \sim \mathcal{N}(\boldsymbol{0},\boldsymbol{\Omega})$, with $\boldsymbol{\Omega}$ denoting the variance-covariance matrix. 

$g^{(1)}$ and $g^{(2)}$ represent the error models, which depend on the additional parameters $\boldsymbol{\xi_1}$ and $\boldsymbol{\xi_2}$, and $\epsilon^{(1)}$ and $\epsilon^{(2)}$ are standard Gaussian variables. The usual error models are the constant model where $g^{(l)}\left(\boldsymbol{\theta_i^{(l)}},t,\xi_l=a\right)=a$, the proportional model where $g^{(l)}\left(\boldsymbol{\theta_i^{(l)}},t,\xi_l=b\right)=bf^{(l)}\left(\boldsymbol{\theta_i^{(l)}},t\right)$ and combinations of the constant and proportional models.

\subsection{PD endpoint toxicity model}

$r(\boldsymbol{\theta_i},\boldsymbol{s_{i,j}}) $ is defined as the function derived from the PK/PD models that returns the value of the PD endpoint (such as the peak of a biomarker) exactly after the administration of the dose-regimen $\boldsymbol{s_{i,j}}$ with individual PK/PD parameters $\boldsymbol{\theta_i}$. Let $\boldsymbol{R(\theta_i,s_{i,j}}) = (r(\boldsymbol{\theta_i},s_{i,1}),...,r(\boldsymbol{\theta_i},\boldsymbol{s_{i,j}}))$ be the function derived from the PK/PD models that returns the vector of all PD endpoints (such as all biomarker peaks) observed after the administration of the regimen $\boldsymbol{s_{i,j}}$ with individual PK/PD parameters $\boldsymbol{\theta_i}$. For patient $i$, we can simplify the notations considering $r_{i,j}=r\left(\boldsymbol{\theta_i},\boldsymbol{s_{i,j}}\right)$, $\boldsymbol{R_{i,j}}=\boldsymbol{R\left(\theta_i,s_{i,j}\right)}$ and the vector of all PD endpoints $\boldsymbol{R_{i}}=\boldsymbol{R_{i,j_i}}$.

Then, let $r^{M}_{i}=\displaystyle\max\limits_{l \in \{1,...,j_i\}}(r_{i,l})$ be the summary PD endpoint (such as the highest peak) observed in patient $i$,which we assume is related to toxicity. 

To define the prior distributions, let $(\overline{r}^M_1,\overline{r}^M_2,...\overline{r}^M_K)$ denote the reference values of the summary endpoint of all dose-regimens of the trial $(\boldsymbol{S_1},...,\boldsymbol{S_k})$; for example we can consider population values $\overline{r}^M_k=\max \left\{ r\left(\boldsymbol{\mu},\boldsymbol{S_{k,1}}\right),...,r\left(\boldsymbol{\mu},\boldsymbol{S_k}\right) \right\}$ with $\boldsymbol{\mu}$ as the PK/PD vector of fixed effects.

In the following section, two statistical models between the PD endpoint and toxicity response are shown.

\subsubsection{Logistic-DRtox}

We propose a Bayesian logistic model to link the global binary toxicity response of patient $i$ receiving $\boldsymbol{s_i}$ to his summary PD endpoint related to toxicity as follows:
\begin{equation}\label{eq:logitmodel}
\text{logit}\left\{\mathbb{P}\left(\displaystyle Y_{i}=1\right)\right\}=\beta_0+\beta_1 \log\left(\displaystyle\frac{r^M_i}{\overline{r}^M_{k_{T}}}\right) 
\end{equation}

where $\beta_1>0$ to have the toxicity probability that increases with the value of the summary PD endpoint. We normalize the PD endpoint for prior elicitation using $\overline{r}^M_{k_{T}}$ which is the reference value of dose-regimen $\boldsymbol{S_{k_T}}$ which we initially guess to have a toxicity probability of $\delta_T$. In this model, we do not consider the longitudinal values of the biomarker as we assume that toxicity is not due to the cumulative effect of the biomarker profile. However, previous drug administrations are considered in the construction of the biomarker through the PK/PD model. Let $\pi_1\left\{\left(\beta_0,\beta_1\right),r^M_i\right\}=\text{logit}^{-1}\left\{\beta_0+\beta_1 \log\left(\displaystyle\frac{r^M_i}{\overline{r}^M_{k_{T}}}\right)\right\}$.

Regarding prior distributions, we consider a normal distribution for the intercept, $\beta_0 \sim \mathcal{N}(\overline{\beta}_0,\sigma_{\beta_0}^2) $ and a gamma distribution for the slope to ensure positivity, $\beta_1 \sim \gamma(\alpha_1,\displaystyle\frac{\alpha_1}{\overline{\beta}_1})$, where $\alpha_1$ is the shape parameter, $\overline{\beta}_0=\mathbb{E}[\beta_0]$ and $\overline{\beta}_1=\mathbb{E}[\beta_1]$.
By construction, we have $\overline{\beta}_0=\text{logit}\left(\delta_T\right)$, obtained via Eq.~\ref{eq:logitmodel} with $r^M_i = \overline{r}^M_{k_{T}}$. Then, let $(p_1,...,p_K)$ be the initial guesses of the toxicity probabilities of regimens $(\boldsymbol{S_1},...,\boldsymbol{S_K})$, where $p_{k_T}=\delta_T$. We can determine $\overline{\beta}_1$ using either only one regimen, which differs from the reference regimen $\boldsymbol{S_{k_T}}$, as  $
\pi_{1}\left\{\left(\overline{\beta_0},\overline{\beta_1}\right),\overline{r}^M_k\right\}=p_k$, with $k \in \{1,...,K\}$ and $k \neq k_T$, or multiple regimens, such as the neighbors of the reference regimen, as follows:
\begin{equation}
\overline{\beta}_1=\argmin_{\beta_1} \sum_{k=k_T-1}^{k_T+1} \left[p_k-\pi_{1}\left\{\left(\overline{\beta}_0,\beta_1\right),\overline{r}^M_k\right\} \right]^2
\end{equation}

\subsubsection{Hierarchical-DRtox}

In this approach, we assume that patients experience toxicity if their PD response exceeds an unknown threshold specific to each patient. To consider inter-individual variability in toxicity, we introduce a patient-specific continuous latent variable, $Z_i$, which represents the toxicity threshold of the PD response. In contrast to the previous approach, we model toxicity after each administration using a modification of the hierarchical probit model as follows \cite{berry_2010}:
\begin{equation}
	\left\{
		\begin{array}{l}
		
		Y_{i,j}=\left\{
			\begin{array}{l}
				0 \quad \text{if } Z_i>\log\left(\displaystyle\frac{r_{i,j}}{\overline{r}^M_{k_{50}}}\right) \\
				1 \quad \text{if } Z_i \leq \log\left(\displaystyle\frac{r_{i,j}}{\overline{r}^M_{k_{50}}}\right) \\
			\end{array}
		\right. \\
		
		Z_i \sim \mathcal{N}(\mu_z,\tau_z^2) \\
		\end{array}
    \right.
    \label{bayesian_hierarchical}
\end{equation}
where $\overline{r}^M_{k_{50}}$ is the reference value at the dose-regimen $\boldsymbol{S_{k_{50}}}$ which we initially guess to have a toxicity probability of 0.5. By adding the random effect, this Bayesian hierarchical model shares common features with the probit model, where $\tau_z^2$ represents the between-subject variance and controls the extent of the borrowing across all patients.

If we consider a new patient $i$ with a vector of biomarker endpoints $\boldsymbol{R_i}$, we can predict his probability of toxicity by $\mathbb{P}\left(Y_i=1\right)=F_z\left\{\displaystyle\log\left(\frac{r^M_{i}}{\overline{r}^M_{k_{50}}}\right)\right\}$, where $F_z$ is the cumulative distribution function of $\mathcal{N}(\mu_z,\tau_z^2)$. The details of the formula are shown in Web Appendix B. Let $\pi_2\left\{\left(\mu_z,\tau_z^2\right),r^M_{i}\right\}=F_z\left\{\displaystyle\log\left(\frac{r^M_{i}}{\overline{r}^M_{k_{50}}}\right)\right\}$.

Regarding the prior distributions, we consider $\mu_z \sim \mathcal{N}(0,\sigma_{\mu_z}^2) $ and $\tau_z \sim \text{half-Cauchy}(0,\sigma_{\tau_z}^2)$. Regarding the half-Cauchy distribution, we followed the recommendations by Gelman, as we assumed that $\tau_z$ could be near 0 \cite{gelman_2006}. Web Appendix F shows how this model can be implemented.

\subsection{Dose-regimen toxicity model}

The posterior toxicity probability of dose-regimen $\boldsymbol{S_k}$ is estimated by integrating the PD endpoint toxicity model on all possible values of the PD endpoint. As this integral cannot usually be solved analytically, the posterior toxicity probability of regimen $\boldsymbol{S_k}$ is estimated via the drawing of an $M$ hypothetical set of patients with M-vector $(p_T(\boldsymbol{S_k})^{(1)},...,p_T(\boldsymbol{S_k})^{(M)})$ as posterior toxicity probabilities. Then, the posterior toxicity probability of regimen $\boldsymbol{S_k}$ is estimated as the posterior mean $\widehat{p_T}(\boldsymbol{S_k})=\displaystyle\frac{1}{M}\sum_{m=1}^{M}p_T(\boldsymbol{S_k})^{(m)}$. This sample of the posterior toxicity probability requires the following three major steps:

\begin{enumerate}
\item \underline{Model fitting:}

\begin{enumerate}
\item First, the PK/PD models are fitted to obtain estimates of the population parameters comprising the fixed effects, $\boldsymbol{\widehat{\mu}}$, and the random effects variance-covariance matrix, $\boldsymbol{\widehat{\Omega}}$, under the Frequentist paradigm. The patients' individual PK/PD parameters, $(\boldsymbol{\widehat{\theta}_1},...,\boldsymbol{\widehat{\theta}_n})$, are also estimated. 
\item Based on the estimated PK/PD parameters, the PD biomarkers are predicted for each patient:

\begin{itemize}
	\item For the logistic-DRtox: the global biomarker peaks $(\widehat{r}^{M}_1,...,\widehat{r}^{M}_n)$ are predicted for each patient as $\widehat{r}^{M}_i= \max \left\{ r\left(\boldsymbol{\widehat{\theta}_i},\boldsymbol{s_{i,1}} \right),...,r\left(\boldsymbol{\widehat{\theta}_i},\boldsymbol{s_i}\right)\right\}$ for $i \in \{1,...,n\}$. 
	The vector of toxicity responses and biomarker responses, $((Y_{1},...,Y_{n}),(\widehat{r}^{M}_1,...,\widehat{r}^{M}_n))$, constitutes the data of the trial.
	
	\item For the hierarchical-DRtox: the biomarker peaks vectors $(\boldsymbol{\widehat{R}_1},...,\boldsymbol{\widehat{R}_n})$ are predicted for each patient  as $\boldsymbol{\widehat{R}_i}=\boldsymbol{R(\widehat{\theta}_i,s_{i})}$ for $i \in \{1,...,n\}$. 
	The vector of toxicity responses and biomarker responses, $((Y_{1,1},...,Y_{n,j_n}),(\boldsymbol{\widehat{R}_1},...,\boldsymbol{\widehat{R}_n}))$, constitutes the data of the trial.
	
\end{itemize}
\item A vector of the parameters of the PD endpoint toxicity model of size $m_\text{iter}$ is sampled from their posterior distribution:

\begin{itemize}
	\item For the logistic-DRtox, $\left(\left(\beta_0^{(1)},\beta_1^{(1)}\right),...,\left(\beta_0^{(m_\text{iter})},\beta_1^{(m_\text{iter})}\right)\right)$ is sampled.
	\item For the hierarchical-DRtox, $\left(\left(\mu_z^{(1)},\tau_z^{(1)}\right),...,\left(\mu_z^{(m_\text{iter})},\tau_z^{(m_\text{iter})}\right)\right)$ is sampled. 
\end{itemize}

\end{enumerate}

\item \underline{Prediction of new patients for the sampling distribution of the PD endpoint:}

\begin{enumerate}

\item The individual PK/PD parameters of $m_\text{predict}$ simulated patients, $\left(\boldsymbol{\theta^{(1)}},...,\boldsymbol{\theta^{(m_\text{predict})}}\right)$, are sampled from $\boldsymbol{\widehat{\mu}}$ and $\boldsymbol{\widehat{\Omega}}$ as $\boldsymbol{\theta^{(m_p)}}=\boldsymbol{\widehat{\mu}} e^{\boldsymbol{\eta^{(m_p)}}}$, with $\boldsymbol{\eta^{(m_p)}} \sim \mathcal{N}(\boldsymbol{0},\boldsymbol{\widehat{\Omega}})$ for $m_p \in \{1,...,m_\text{predict}\}$

\item The maximum biomarker endpoint of each simulated patient receiving regimen $\boldsymbol{S_k}$ is obtained as $r^{M^{(m_p)}}=\max \left( r\left(\boldsymbol{\theta^{(m_p)}},\boldsymbol{S_{k,1}}\right),...,r\left(\boldsymbol{\theta^{(m_p)}},\boldsymbol{S_k}\right) \right)$ for $m_p \in \{1,...,m_\text{predict}\}$
\end{enumerate}

\item \underline{Estimation of the posterior distribution of the probability of toxicity:}
\begin{enumerate}

\item The $m^\text{th}$ iteration, $m=(m_i,m_p) \in \{1,...,M\}$, where $M=m_\text{iter}*m_\text{predict}$, of the posterior probability of toxicity of dose-regimen $\boldsymbol{S_k}$, $p_T(\boldsymbol{S_k})^{(m)}$, is obtained depending on the method chosen:

\begin{itemize}
	\item For the logistic-DRtox, $p_T(\boldsymbol{S_k})^{(m)}=\pi_1\left\{\left(\beta_0^{(m_i)},\beta_1^{(m_i)}\right),r^{M^{(m_p)}}\right\}$
	
	\item For the hierarchical-DRtox, $p_T(\boldsymbol{S_k})^{(m)}=\pi_2\left\{\left(\mu_z^{(m_i)},\tau_z^{(m_i)}\right),r^{M^{(m_p)}}\right\}$ \\
\end{itemize}

\end{enumerate}

\end{enumerate}

The DRtox approach allows us to estimate the toxicity probability of the panel of dose-regimens $\boldsymbol{\mathcal{S}}$ and predict the toxicity probability of each new regimen defined from the set of doses $\boldsymbol{\mathcal{D}}$.

\section{Simulation study}
\label{simu_study}

\subsection{Simulation settings}

The performance of the DRtox was evaluated through a simulation study. We assumed that toxicity was related to a PD endpoint (the peak of cytokine in the context of our motivating example). Therefore, to simulate toxicity, we first needed to simulate the PK/PD profiles and simulate toxicity from the PD profile.

Regarding the PK/PD models, we were inspired by published models on blinatumomab, which is another bispecific T-cell engager that binds to CD3 on T-cells and to CD19 on tumor cells. Regarding the PK model, we considered a 1-compartment infusion model in which the parameters are the volume of distribution V and the clearance of elimination Cl and assumed 4 hours of infusion \cite{zhu_2016}. The model is defined in Web Appendix A. Regarding the PD aspect, the objective was to model cytokine mitigation in the case of intra-patient dose-escalation. We simplified the model developed by Chen et al., which assumes that cytokine production is stimulated by the drug concentration but inhibited by cytokine exposure through the AUC \cite{chen_2019}. We defined the PD model as follows: 
\begin{equation}
\displaystyle\frac{\mathrm{d} E\left(t\right)}{\mathrm{d} t}=\displaystyle\frac{E_{max}C\left(t\right)^H}{{EC_{50}}^H+C\left(t\right)^H}\left \{ 1-\displaystyle\frac{I_{max}AUC_E\left(t\right)}{\displaystyle\frac{IC_{50}}{K^{J-1}}+AUC_E\left(t\right)} \right \}-k_{deg}E\left(t\right)
\end{equation}
 where $E\left(t\right)$ and $C\left(t\right)$ are the cytokine and drug concentration at time t, respectively, $AUC_E\left(t\right)$ is the cumulative cytokine exposure, and the parameters are defined in Table \ref{table_PKPD_param}. Additional information concerning the PK/PD models is provided in Web Appendix A.

In both the PK and PD models, we considered a proportional error model with $b=0.1$. The values of the PK/PD parameters used for the simulations were inspired by the estimated parameters of blinatumomab and are displayed in Table \ref{table_PKPD_param} \cite{zhu_2016}\cite{chen_2019}.

\begin{table}[h!]
\caption{Definition and values of the PK/PD parameters used for the simulation study. Parameter estimates represent the fixed effects, and coefficients of variation (CV) are the square root of the diagonal of the variance-covariance matrix. They are inspired by the parameters estimated on blinatumomab, with a modification of I$_\text{max}$ to observe cytokine mitigation after several administrations \cite{zhu_2016}\cite{chen_2019}.
\label{table_PKPD_param}}
\begin{center}
\begin{tabular}{c|cccc}
 & \multirow{2}{*}{Parameter} & \multirow{2}{*}{\shortstack{Estimate\\\\ (\% CV)}} & \multirow{2}{*}{Unit} & \multirow{2}{*}{Description}   \\ \\ \hline
\multirow{2}{*}{ \shortstack{PK\\\\model}} & Cl & $1.36$ $(41.9)$ & L/h & Clearance \\
& V & $3.4$ $(0)$ & L & Distribution volume \\ \hline
\multirow{7}{*}{\shortstack{PD\\\\model}} & E$_\text{max}$ & $3.59 \cdot 10^5$ $(14)$ & pg/mL/h & Max cytokine release rate \\
& EC$_\text{50}$ & $1 \cdot 10^4$ $(0)$ & ng/mL & Drug exposure for half-max release \\
& H & $0.92$ $(3)$ &       &Hill coefficient for cytokine release \\ 
& I$_\text{max}$ & $0.995$ $(0)$     &      & Max inhibition of cytokine relase \\
& IC$_\text{50}$  & $1.82 \cdot 10^4$  $(12)$    & pg/mL$\cdot$h & Cytokine exposure for half-max inhibition \\
& k$_\text{deg}$ & $0.18$ $(13)$      & h$^{-1}$ & Degradation rate for cytokine  \\
& K & $2.83$ $(36)$         &   & Priming factor for cytokine relase\\ 
\end{tabular}
\end{center}
\end{table}

To simplify and accelerate the PK/PD estimation during the simulations, we followed the traditional PK/PD modeling strategy for small sample size data by fixing some parameters. We considered the parameters EC$_\text{50}$, I$_\text{max}$ and IC$_\text{50}$ fixed and no random effects on V and H. Our modeling choices can be challenged in practice, but the objective of this work is not to propose a PK/PD model but rather to propose a global modeling approach including PK/PD estimation in a phase I toxicity model. In our case, we decided to use a previously validated PK/PD model that mimics the behavior we expect in our motivating trial to demonstrate the performance of the proposed framework.

\begin{figure}[h!]
\begin{center}
\centerline{\includegraphics[width=16cm]{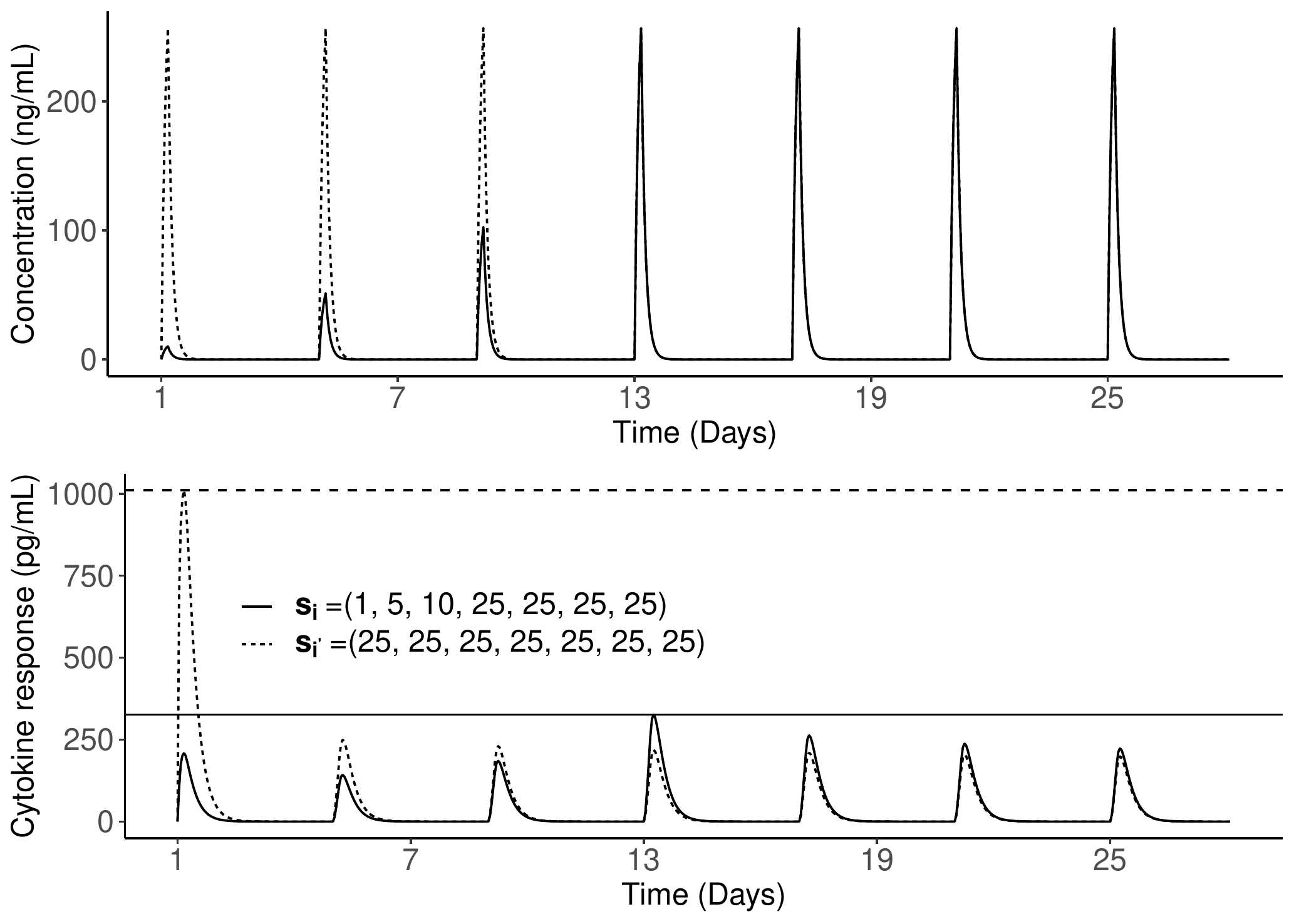}}
\end{center}
\caption{Concentration (up) and cytokine (down) profiles of two patients, one receiving a dose-regimen with intra-patient escalation in solid line and the other receiving a dose-regimen without intra-patient escalation in dashed line, administered on days 1, 5, 9, 13, 17, 21 and 25. Horizontal lines represent the maximum peak of cytokine observed after each dose-regimen.
\label{plot_PKPD}}
\end{figure}

We used as the PD endpoint $r_{i,j}$ the peak of cytokine observed for patient $i$ after the $j^\text{th}$ administration, and for $r^{M}_{i}$ the highest peak of cytokine observed for patient $i$. Using the PK/PD models presented above and the parameters shown in Table \ref{table_PKPD_param}, we were able to model the mitigation of cytokine release upon repeating dosing, which was reflected by the decrease in the cytokine peak with repeating dosing. Hence, we were able to model that slowly increasing the dose reduces the cytokine peak compared to directly giving the steady-state dose.
For example, we compared the concentration and cytokine profiles of patients $i$ and $i'$ who received regimens $\boldsymbol{s_i}=(1,5,10,25,25,25,25)$ $\mu$g/kg and $\boldsymbol{s_{i'}}=(25,25,25,25,25,25,25)$ $\mu$g/kg administered on days 1, 5, 9, 13, 17, 21 and 25 (Figure \ref{plot_PKPD}). From the $4^\text{th}$ administration, the concentration profiles of patients $i$ and $i'$ are the same, but in the cytokine profile, the maximum peak of cytokine of patient $i'$ is much higher than that of patient $i$, $r^{M}_{i'}=r_{i',1}>r^{M}_{i}=r_{i,4}$. 

To simulate toxicity from the cytokine profile, we defined a threshold $\tau_{\scriptscriptstyle{T}}$ on the cytokine response and assumed that toxicity occurred if this threshold was exceeded \cite{ursino_2017}. To introduce between-subject variability, we defined a log-normally distributed measure of subject sensitivity, $\alpha_i$ for patient $i$, where $\alpha_i=\text{e}^{\eta_{\alpha_i}}$ and  $\eta_{\alpha_i} \sim \mathcal{N}(0,\omega_\alpha^2)$.
We assumed that patient $i$ experienced toxicity at the $j^{\text{th}}$ administration, $Y_{i,j}=1$, if $\alpha_i r_{i,j} \geq \tau_{\scriptscriptstyle{T}}$.

To compute the toxicity probability of regimen $\boldsymbol{S_k}$, we used the Monte-Carlo method by simulating $N=10000$ cytokine profiles under $\boldsymbol{S_k}$ and computing
\begin{equation}
p_T(\boldsymbol{S_k}) = \displaystyle\frac{1}{N} \sum_{i=1}^{N} 
\left[ 
1-\Phi\left\{
\frac{\text{log}\left(\tau_{\scriptscriptstyle{T}}\right)-\text{log}\left(r_i^M\right)}{\omega_\alpha}
\right\}
\right]
\end{equation}
where $\Phi$ is the cumulative distribution function of the standard normal distribution. 

\begin{figure}
\begin{center}
\centerline{\includegraphics[width=16cm]{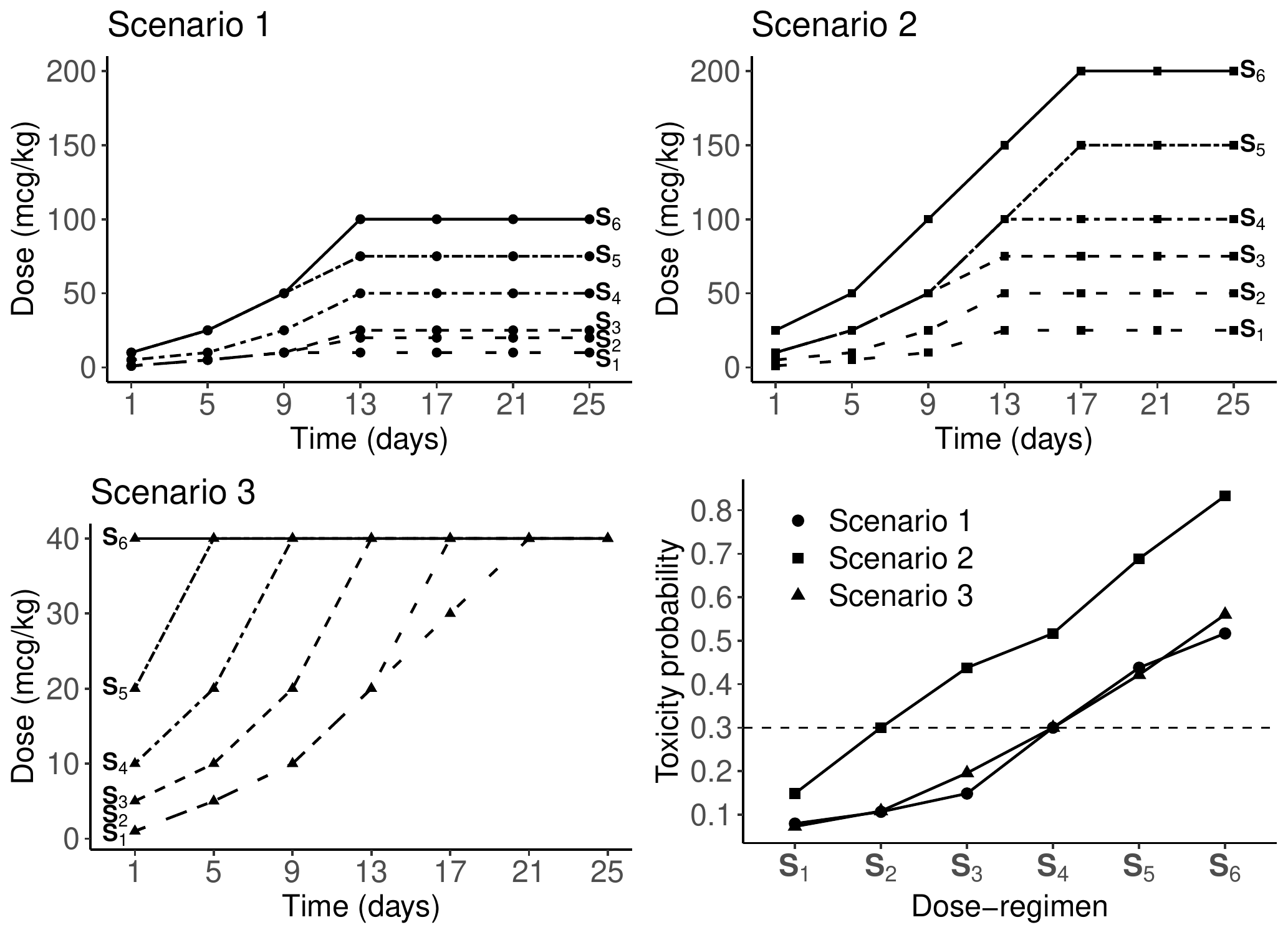}}
\end{center}
\caption{The first 3 subplots represent the panel of dose-regimens from $\boldsymbol{S_1}$ in spaced dashed line to $\boldsymbol{S_6}$ in solid line, for the 3 main scenarios, where the type of points is specific to each scenario. In the last subplot in the lower right corner, the dose-regimen toxicity relationship is represented for each scenario, where the MTD-regimen is the dose-regimen having the toxicity probability the closest to the target $\delta_T$, plotted in dashed line.
\label{plot_scenarios_1_3}}
\end{figure}

We present the results of 3 toxicity scenarios by varying the dose-regimens and the value of the threshold $\tau_{\scriptscriptstyle{T}}$ to explore different positions of the MTD-regimen (with $\omega_\alpha=0.25$). Additional scenarios are presented in Web Appendix D. In each scenario, we considered 6 dose-regimens, and each dose-regimen included 7 dose-administrations on days 1, 5, 9, 13, 17, 21 and 25. The dose-regimens chosen for each scenario and the dose-regimen toxicity curves are displayed in Figure \ref{plot_scenarios_1_3}. In Scenarios $1$, $2$ and $3$, the MTD-regimen is situated at dose-regimens $\boldsymbol{S_4}$, $\boldsymbol{S_2}$ and $\boldsymbol{S_4}$, respectively. Scenarios $1$ and $2$ are inspired from the motivating trial in which the dose-regimens reach the steady-state dose at approximately the same time, and have increasing steady-state doses. However, Scenario $3$ represents a case in which the objective is to reach the steady-state dose, 40 $\mu$g/kg, as fast as possible to increase potential efficacy under toxicity constraints. The dose-regimen toxicity relationship is similar to that in Scenario 1 but with less difference between the MTD-regimen and its neighbors.

For each scenario, 1000 trials were simulated, and $\delta_{T}=0.3$ was considered the toxicity target. Because we applied our methods once all patients from the trial were included, we evaluated the impact of 2 traditional dose-escalation designs, i.e., the 3+3 design and a modified continual reassessment method (CRM) initially proposed by O'Quigley et al. \cite{oquigley_1990}. A flow diagram of the rules of the 3+3 design is provided in Web Appendix E. For the modified CRM, we considered a 2-parameter logistic regression model with cohorts of a size of 3 and a total sample size of 30 patients \cite{cheung_2011}. Dose-skipping was not allowed, and early stopping rules were not implemented. We based the skeleton of the CRM, i.e., the prior guesses of the toxicity probabilities, on Scenario 1, i.e., $(0.06, 0.12, 0.20, 0.30, 0.40, 0.50)$. This skeleton was used in all simulations and scenarios.

When defining the prior distributions for our proposed models, we calibrated the model prior distributions based on the initial guesses of the toxicity probabilities (we used the same initial guesses for the CRM). To quantify the information provided by the prior distribution, we computed the approximate effective sample size (ESS), which was defined as the equivalent sample size embedded in the prior distribution of the model parameters \cite{yuan_2017}. In practice, we approximated the ESS by matching the mean and variance of the toxicity probabilities computed from the prior distributions to those of a beta distribution. Then, the ESS was computed as the sum of the parameters of the beta distribution. More details of the ESS computation are shown in Web Appendix C. In our settings, for the logistic-DRtox, we considered $k_T=4$, $\sigma_{\beta_0}=2$ and $\alpha=5$, leading to an approximate mean ESS of $1.6$. For the hierarchical-DRtox, we considered $k_{50}=6$, $\sigma_{\mu}=1$ and $\sigma_{\tau}=1$, leading to an approximate mean ESS of 1.8.

All simulations were performed in the R environment, using Monolix software for the PK/PD estimation and Stan for the Bayesian analysis \cite{r_2018}\cite{monolix_2019}\cite{stan_2019}. 

\subsection{Simulation results}

\subsubsection{Proportions of correct selection}
We first evaluated the performance of the DRtox according to the proportions of correct selection (PCS) based on the proportions that each regimen is selected as the MTD-regimen over the trials. We evaluated the impact of the dose-regimens and the position of the MTD-regimen in 3 toxicity scenarios, and the impact of the dose-escalation design, i.e., either the 3+3 design or the CRM. The PCS results of the 3 main scenarios and the mean sample size of each dose-regimen across the trials due to the chosen dose-escalation design are displayed in Table \ref{table_PCS}. The PCS of additional scenarios are displayed in Web Appendix D. As a practical rule, we could only recommend as the MTD-regimen a dose-regimen that was administered during the dose-escalation phase of the trial.

\begin{table}[]
\caption{Proportions that each dose-regimen is being selected as the MTD-regimen over the 1000 trials in the 3 toxicity scenarios and the 2 dose-allocation designs, either the 3+3 design or the CRM. For each scenario, the PCS on the true MTD-regimen are represented in bold. For each dose-allocation design, the mean sample size of each dose-regimen is displayed.
\label{table_PCS}}
\centering
\begin{tabular}{llllllll}
\cline{3-8}
                                                                                           &        & $\boldsymbol{S_1}$ & $\boldsymbol{S_2}$ & $\boldsymbol{S_3}$ & $\boldsymbol{S_4}$ & $\boldsymbol{S_5}$ & $\boldsymbol{S_6}$ \\ \hline
\multicolumn{2}{l}{\textbf{Scenario 1}}                                                                   &  \textbf{0.08} & \textbf{0.11} & \textbf{0.15} & \textbf{0.3} & \textbf{0.44} & \textbf{0.52}   \\ \cline{1-8} 
\multicolumn{1}{l}{\multirow{4}{*}{3+3}} & Mean sample size & 3.6 & 3.5 & 3.5 & \textbf{3} & 1.6 & 0.4 \\ \cline{2-8} 
\multicolumn{1}{l}{}                     & Logistic-DRtox   &  8.6 & 5.9 & 19 & \textbf{42.2} & 19.6 & 4.7 \\
\multicolumn{1}{l}{}                     & Hierarchical-DRtox   &  7.5 & 7.6 & 19.1 & \textbf{43.8} & 18.6 & 3.4 \\
\multicolumn{1}{l}{}                     & 3+3    & 13.9 & 16.1 & 32.2 & \textbf{27.6} & 8.6 & 1.6 \\ \cline{1-8} 
\multicolumn{1}{l}{\multirow{4}{*}{CRM}} & Mean sample size &  4.2 & 3.7 & 5.6 & \textbf{8.8} & 5.6 & 2.1 \\ \cline{2-8} 
\multicolumn{1}{l}{}                     & Logistic-DRtox   & 0 & 1.2 & 15.5 & \textbf{64.6} & 15.5 & 3.2 \\
\multicolumn{1}{l}{}                     & Hierarchical-DRtox   &  0 & 0.8 & 12.8 & \textbf{64.3} & 19.4 & 2.7 \\
\multicolumn{1}{l}{}                     & Logistic CRM    &   0 & 1.4 & 15.1 & \textbf{50.4} & 27.1 & 6  \\ \cline{1-8}

\multicolumn{2}{l}{\textbf{Scenario 2}}                                                                   &   \textbf{0.15} & \textbf{0.3} & \textbf{0.44} & \textbf{0.52} & \textbf{0.69} & \textbf{0.83}   \\ \cline{1-8} 
\multicolumn{1}{l}{\multirow{4}{*}{3+3}} & Mean sample size &4 & \textbf{3.6} & 1.8 & 0.5 & 0.1 & 0\\ \cline{2-8} 
\multicolumn{1}{l}{}                     & Logistic-DRtox   &  27.2 & \textbf{42.5} & 24.7 & 5.2 & 0.4 & 0 \\
\multicolumn{1}{l}{}                     & Hierarchical-DRtox   &  29.3 & \textbf{41.2} & 24.3 & 4.8 & 0.4 & 0 \\
\multicolumn{1}{l}{}                     & 3+3    & 57.3 & \textbf{31} & 9.8 & 1.7 & 0.2 & 0 \\ \cline{1-8} 
\multicolumn{1}{l}{\multirow{4}{*}{CRM}} & Mean sample size &  8.7 & \textbf{11.1} & 7.5 & 2.3 & 0.3 & 0 \\ \cline{2-8} 
\multicolumn{1}{l}{}                     & Logistic-DRtox   & 14.8 & \textbf{65.9} & 17.4 & 1.7 & 0.2 & 0 \\
\multicolumn{1}{l}{}                     & Hierarchical-DRtox   &  12.3 & \textbf{66.2} & 18.9 & 2.6 & 0 & 0 \\
\multicolumn{1}{l}{}                     & Logistic CRM    &  12.5 & \textbf{56} & 26.7 & 4.7 & 0.1 & 0  \\ \cline{1-8}

\multicolumn{2}{l}{\textbf{Scenario 3}}                                                                   &  \textbf{0.07} & \textbf{0.11} & \textbf{0.2} & \textbf{0.3} & \textbf{0.42} & \textbf{0.56}   \\ \cline{1-8} 
\multicolumn{1}{l}{\multirow{4}{*}{3+3}} & Mean sample size & 3.6 & 3.6 & 3.7 & \textbf{2.7} & 1.4 & 0.4 \\ \cline{2-8} 
\multicolumn{1}{l}{}                     & Logistic-DRtox   &  7.8 & 6.4 & 25.2 & \textbf{34.1} & 21.6 & 4.9 \\
\multicolumn{1}{l}{}                     & Hierarchical-DRtox   &  5.9 & 7.9 & 27.3 & \textbf{35.8} & 20.6 & 2.5\\
\multicolumn{1}{l}{}                     & 3+3    & 13.1 & 24.4 & 29.5 & \textbf{24} & 7.7 & 1.3 \\ \cline{1-8} 
\multicolumn{1}{l}{\multirow{4}{*}{CRM}} & Mean sample size &  4 & 4 & 6.4 & \textbf{8} & 5.2 & 2.3 \\ \cline{2-8} 
\multicolumn{1}{l}{}                     & Logistic-DRtox   & 0.1 & 1.4 & 19.6 & \textbf{52} & 25.1 & 1.8 \\
\multicolumn{1}{l}{}                     & Hierarchical-DRtox   & 0.1 & 0.8 & 17.7 & \textbf{54.4} & 25.9 & 1.1 \\
\multicolumn{1}{l}{}                     & Logistic CRM    &   0.1 & 2.3 & 20.3 & \textbf{44.5} & 26.4 & 6.4  \\ \cline{1-8}
\end{tabular}
\end{table}

In all scenarios, the PCS of the logistic-DRtox and the hierarchical-DRtox are very similar. Both methods outperform the dose-escalation design implemented in most scenarios. After implementing the 3+3 design, our methods correctly select the MTD-regimen in more than 10\% more trials compared to the dose-allocation design. After implementing the CRM design, both methods correctly select the MTD-regimen in approximately 10\% more trials compared to the CRM. 

The results of Scenarios 1, 3 and 6 (presented in Web Appendix D) illustrate the effect of the variation in the dose-regimen scheme with a similar dose-regimen toxicity relationship. Compared to Scenario 1, the PCS of the logistic-DRtox and hierarchical-DRtox are decreased by approximately 10\% in Scenario 3, while there is not much difference in the results between Scenarios 1 and 6. Therefore, the loss of performance in Scenario 3 is caused not only by the variation in the dose-regimen scheme but also by the difference in the dose-regimen toxicity relationship, as in Scenario 3 there is less difference in the toxicity probabilities between the MTD-regimen and its neighbors.

However, the performance of the DRtox is heavily impacted by the dose-escalation design implemented; after implementing the CRM design, the DRtox correctly selects the MTD-regimen in more than $50 \%$ of the trials, but its PCS can decrease by $20 \%$ when applied after the 3+3 design. This loss of performance is due to the small sample size after implementing the 3+3 design and the higher proportion of patients allocated to suboptimal dose-regimens.

\subsubsection{Estimation of the toxicity probabilities}

We also evaluated the performance of the DRtox based on the precision of the estimation of the toxicity probabilities of all dose-regimens. We represented the distribution of the estimated toxicity probabilities, defined as the mean of the posterior distribution, over 1000 trials. The results of Scenario 3 obtained after implementing the CRM are presented in the lower part of Figure \ref{plot_pTox_predict}. The results of the other scenarios are displayed in Web Appendix D.

In all scenarios, the toxicity probability of the MTD-regimen is well estimated by the DRtox and the CRM. Both the hierarchical-DRtox and the logistic-DRtox seem to be better in estimating the toxicity probability at all dose-regimens, even those far from the MTD-regimen. This phenomenon could be due to the additional PK/PD information and the correct understanding of the toxicity mechanism. Using the CRM, the entire dose-regimen toxicity curve is well estimated when the skeleton is close to the truth, as shown in Scenario 1 (Web Appendix D). However, in most cases, the toxicity estimation is precise around the MTD-regimen, but not reliable for the other dose-regimens. Regarding the dose-regimens far from the MTD-regimen, the hierarchical-DRtox seems to estimate the toxicity probability with less bias but more variance than the logistic-DRtox. In Web Appendix D, the distribution of the root mean square error (RMSE) of all methods is plotted; the RMSE is computed on all dose-regimens or on the MTD-regimen and its neighbors. Near the MTD-regimen, the estimation of the logistic-DRtox is better than that of the hierarchical-DRtox; both models are better than the CRM. However, in the scenarios in which the MTD-regimen is at extreme positions of the dose-regimens panel (Scenarios 2 and 4 in Web Appendix D), the entire dose-regimen toxicity relationship is better estimated with the hierarchical-DRtox than the logistic-DRtox.

\subsubsection{Recommendation of a more suitable untested dose-regimen}

\begin{figure}
\begin{center}
\centerline{\includegraphics[width=15cm]{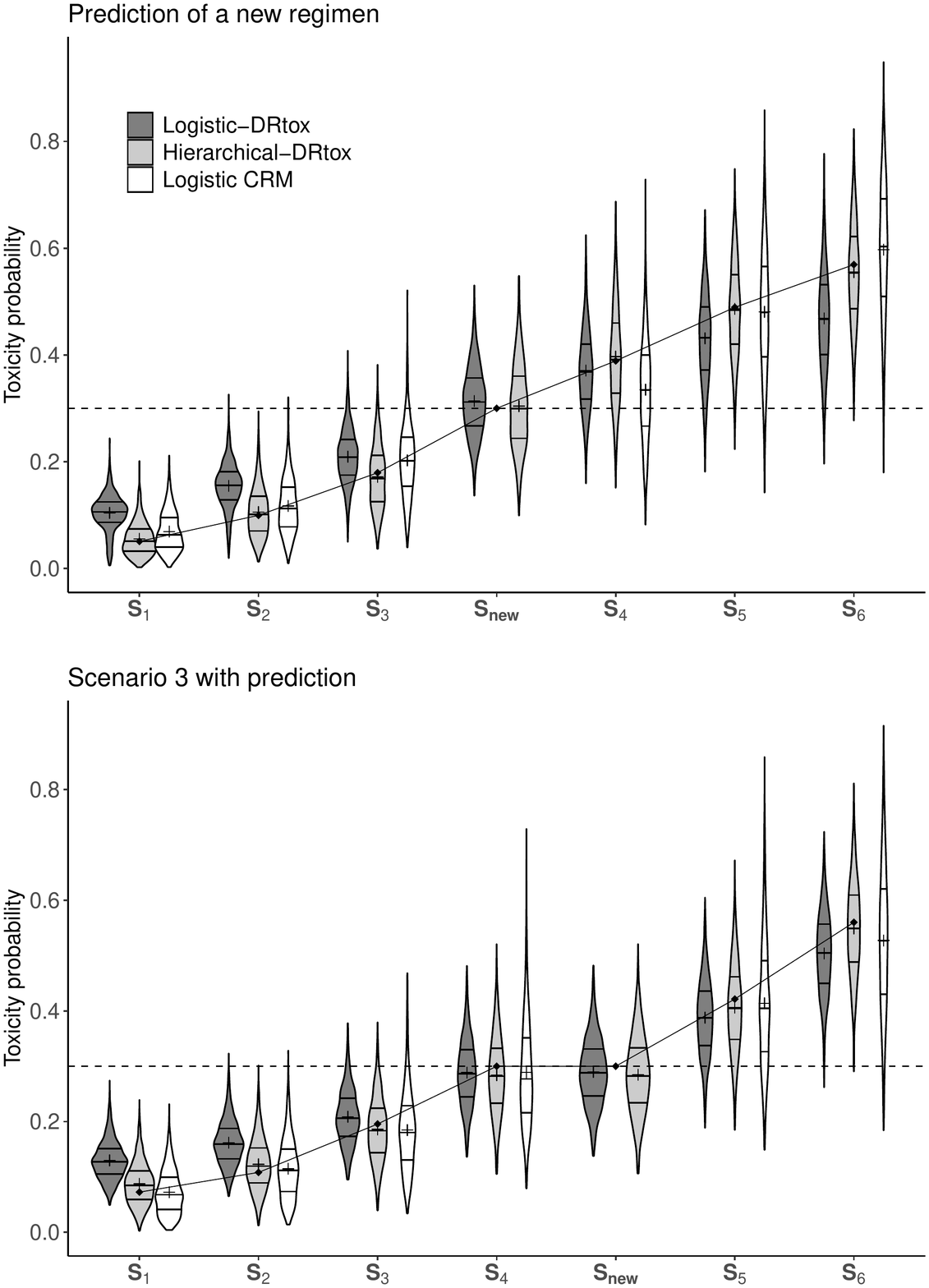}}
\end{center}
\caption{Violin plots of the estimated toxicity probabilities in an additional scenario in which the dose-regimen panel missed the true MTD-regimen and in Scenario 3 on 1000 trials implemented with the CRM including 30 patients. The predicted toxicity probability of a new regimen $\boldsymbol{S_\text{new}}$ is framed in dotted line. Horizontal lines on the density estimates represent the median and first and third quantiles of the distributions and the plus sign represents the mean. The dashed line represents the toxicity target and the solid line represents the true toxicity probabilities.
\label{plot_pTox_predict}}
\end{figure}

Finally, one strength of the DRtox is that it models the entire relationship between the dose-regimen and toxicity and can predict the toxicity probability of any new dose-regimen. Notably, in this work, we assumed that the administration times were fixed to simplify the notations but regimens with different times of drug administration can also be considered. Therefore, at the end of the dose-escalation stage of the trial, the DRtox can recommend dose-regimens that were not tested in the trial to be investigated in expansion studies. For example, let us imagine a scenario in which the panel of dose-regimens missed the true MTD-regimen, as illustrated in the upper plot of Figure \ref{plot_pTox_predict}, where regimen $\boldsymbol{S_3}=(5,10,25,50,50,50,50)$ $\mu$g/kg is under-dosing and regimen $\boldsymbol{S_4}=(10,25,50,100,150,150,150)$ $\mu$g/kg is overdosing.

The upper plot of Figure \ref{plot_pTox_predict} illustrates the gap between the estimated toxicity probabilities of regimens $\boldsymbol{S_3}$ and $\boldsymbol{S_4}$, suggesting that an alternative regimen could be found to have a toxicity probability closer to the target. At the end of the dose-escalation stage of the trial, the DRtox can predict the toxicity probability of any new regimen, such as regimen $\boldsymbol{S_\text{new}}=(10,25,50,100,100,100,100)$ $\mu$g/kg, whereas the CRM is unable to perform predictions as the model is built on a skeleton based on the panel of dose-regimens. In the upper plot of Figure \ref{plot_pTox_predict}, we can observe that both the hierarchical-DRtox and the logistic-DRtox predict that new regimen $\boldsymbol{S_\text{new}}$ has a toxicity probability closer to the target; therefore we can propose to evaluate the new regimen in expansion cohorts.

Another practical case is illustrated in Scenario 3 in which the objective was to administer the steady-state dose of $40$ $\mu$g/kg as soon as possible. As shown in the lower plot of Figure \ref{plot_pTox_predict}, the estimated MTD-regimen is $\boldsymbol{S_4}=(10, 20, 40, 40, 40, 40, 40)$ $\mu$g/kg, and the next regimen of the panel, $\boldsymbol{S_5}=(20, 40, 40, 40, 40, 40, 40)$, is estimated to be too toxic. Nevertheless, one might wonder whether another regimen with an acceptable toxicity could be found in which the steady-state dose is administered from the second administration. The DRtox predicts the toxicity probability of new regimen $\boldsymbol{S_\text{new}}=(10, 40, 40, 40, 40, 40, 40) $ to be approximately 0.3 as shown in the lower plot of Figure \ref{plot_pTox_predict}, and this new regimen can be compared in terms of efficacy to the estimated MTD-regimen $\boldsymbol{S_4}$ in subsequent stages of the trial. Therefore, at the end of the trial, the DRtox can evaluate alternative regimens that were not included in the panel for future studies.

\subsubsection{Sensitivity analysis}

We also evaluated the DRtox under different prior distributions and after increasing the variability in toxicity. The results of these analyses are shown in Web Appendix D. 

To evaluate the impact of the prior distributions, we compared the main results with those obtained with a stronger prior distribution measured with an approximate ESS of 9, which is high for a trial including 30 patients. As the prior distributions are based on Scenario 1, stronger prior information increased the performance in the scenarios in which the dose-regimen toxicity relationship is similar to that in Scenario 1 (Scenarios 1, 3, 5 and 6), but the performance was decreased in Scenario 4. Therefore, defining prior distributions using reliable data from previous studies can increase the performance, but attention should be paid to the quality and quantity of the information used to avoid decreasing the performance. 

We also observed that our methods were robust to an increase in the variability in toxicity by increasing $\omega_\alpha$ from $0.25$ to $0.5$ while maintaining the PK/PD variability unchanged.

\section{Discussion}
\label{s:discuss}

In this work, we developed a dose-regimen assessment method (DRtox) to model the relationship between the dose-regimen and toxicity by modeling a PD endpoint. We estimated the toxicity related to the PD endpoint in the context of an ongoing phase I trial in which the assumption of a monotonic increase in the dose-regimen toxicity relationship did not hold. We found that when the process generating toxicity was reasonably understood and approximated, adding PK/PD information increased the proportion of correct selection (PCS). This method allowed for a better estimation of the dose-regimen toxicity curves, as this type of modeling enabled the sharing of more information across regimens. Moreover, the DRtox was able to evaluate additional regimens for expansion cohorts that were not present in the dose-regimen panel set but may have a predicted toxicity probability closer to the target. In practice, our methods should be applied at the end of the dose-escalation phase of the motivating trial once all PK/PD and toxicity data are collected. Our model can address missing data as follows; (1) Regarding missing doses in the dose-regimen and associated cytokine profiles, as we are using nonlinear PK/PD modeling, our method would take into account whether a patient misses one or more planned doses as the model considers the actual regimen received and not the planned regimen. (2) Regarding missing cytokine data, which is expected to be rare in this trial as the cytokine is carefully monitored by frequent sampling to detect its peak, individual cytokine peaks could be predicted from the population PK/PD model. However, it would be more common for PK/PD data to be below the limit of quantification, but these data are considered by the PK/PD model as censored data rather than missing data. (3) Regarding missing toxicity data, patients with missing data should be replaced.

In the simulation study, we assumed that the dose-regimens were ordered, but the DRtox can be applied when only partial ordering is known. As the DRtox is applied at the end of the trial, the choice of the dose-escalation design may have a significant impact on the results. The performance achieved using a model-based design, such as the CRM with 30 patients, is better than that achieved using an algorithm-based design, such as the 3+3 design, which has the main disadvantage of treating most patients at subtherapeutic doses and having a small total sample size that cannot be fixed before the trial.

Regarding the logistic-DRtox, since drug administration is stopped in the case of toxicity, the performance can be impacted by incomplete observations of the PD endpoint, even though it seemed to lightly impact our simulation study. In the case toxicities occur at the beginning of the administrations, resulting in a high number of incomplete PD observations, we propose the use of the predicted PD given by the PK/PD model under the complete regimen planned. 

The hierarchical-DRtox added a constraint, i.e., toxicity must occur at the maximum of the PD response. Errors in the PK/PD estimation may lead to an undefined hierarchical model. In our simulation study, we observed this latter issue in less than $2 \%$ of the trials. In the real world, this issue could indicate that the proposed PK/PD model is incorrect, and that another model should be considered. However, in our simulation study, we decided to run other simulated datasets to replace the upper $2 \%$ of the trials for all methods. One way to relax this constraint is to allow the patients' toxicity threshold to vary among administrations by adding a second latent variable, which could lead to complex models that are challenging to estimate.

In this work, we assumed that all dose-regimens have the same repetition scheme and duration. However, the DRtox can address regimens with different schemes, administration modes, etc. The first part of the DRtox relies on PK/PD modeling. An incorrect PK/PD model can have a negative impact on the full method. However, as usual in the PK/PD field, the aim of the model is to well catch the outcome profiles; therefore an "incorrect/approximated" model could still be applied without DRtox performance loss.
In conclusion, we proposed a general approach for modeling toxicity through a PK/PD endpoint. In this work, we considered a specific PD endpoint in the context of an actual ongoing clinical trial, but various endpoints (such as the AUC or a combination of several toxicity biomarkers) could be used depending on the type of toxicity considered. Moreover, we developed the DRtox under the assumption that toxicity was linked to the maximum value of the PD biomarker, but other assumptions could be raised, such as assuming a cumulative effect. The usual dose-finding designs were developed to determine the MTD in the first cycle of treatment after a single administration. However, with the increase in the number of targeted molecules, immuno-oncology therapies and combinations with alternative dose-regimens, standard dose-allocation designs fail to identify the dose-regimen recommended for future studies. Incorporating PK/PD exposure data in early phase toxicity modeling through stronger collaboration between biostatisticians and pharmacometricians may lead to a better understanding of the entire dose-regimen toxicity relationship and provide alternative dosage recommendation for the next phases of the clinical development.

\section*{Acknowledgements}
This work was partially funded by a grant from the Association Nationale de la Recherche et de la Technologie, with Sanofi-Aventis R\&D, Convention industrielle de formation par la recherche number 2018/0530.

The authors would like to thank Raouf El-Cheikh, Laurent Nguyen and Christine Veyrat-Follet for their time and explanations of PK/PD modeling and Paula Fraenkel for her thoughtful review of the manuscript.

\end{document}